# Slow light with a swept-frequency source

**Rui Zhang,\* Yunhui Zhu, Jing Wang, and Daniel J. Gauthier**

*Department of Physics, Duke University, Durham, North Carolina, 27708, USA*
*\*rz10@phy.duke.edu*

**Abstra**ct: We introduce a new concept for stimulated-Brillouin-scattering-based slow light in optical fibers that is applicable for broadly-tunable frequency-swept sources. It allows slow light to be achieved, in principle, over the entire transparency window of the optical fiber. We demonstrate a slow light delay of 10 ns at 1.55 μm using a 10-m-long photonic crystal fiber with a source sweep rate of 400 MHz/μs and a pump power of 200 mW. We also show that there exists a maximal delay obtainable by this method, which is set by the SBS threshold, independent of sweep rate. For our fiber with optimum length, this maximum delay is ~38 ns, obtained for a pump power of 760 mW.

## 1. Introduction

Slow light, where the group velocity of a pulse is much smaller than the light speed in vacuum [1,2], has shown promise in a great number of applications [3-5], such as optical buffering, variable true time delay, and enhancing nonlinear optical phenomena. Slow light via stimulated Brillouin scattering (SBS) in optical fibers has attracted a lot of interest due to its compatibility with fiber-optic communication systems and room-temperature operation [6-9]. Its use in high-bandwidth systems is limited, however, due to the natural linewidth of the SBS process, which is typically 35 MHz in conventional single-mode fibers. The bandwidth can be extended into the tens of GHz range by dithering the frequency of the pump beam [10-13]. In some applications, such as swept-source optical coherence tomography (OCT) [14] and Fourier transform spectroscopy [15], it is desirable to have controllable slow light over a bandwidth of several nanometers, which is currently not accessible with SBS slow light.

Optical coherence tomographic systems, comprised of a reference arm and a sample arm, are used for micron-resolution imaging. One method, which shows advantage for obtaining a high signal-to-noise ratio, uses a rapidly-swept narrowband source that is linearly chirped over a broad optical bandwidth. The resolution of such a swept-source OCT system is limited by the total sweep range, and the imaging depth is limited by the smallest frequency step of the source or the digital sampling rate of the detection system. To capture an image of a sample over a large depth, the length of the reference is often changed and the sweep repeated, thereby changing the region of interest. A controllable slow light medium, placed in the reference arm, could make it possible to change the region of interest without changing the physical length of the reference arm because the image location is proportional to group index $n_g$ in the reference arm [16]. However, due to the narrow linewidth of SBS process, it is not possible to use SBS slow light for such an application.

In this paper, we introduced a new concept to realize SBS slow light applicable to broadly-swept sources. This method allows slow light to be achieved, in principle, over the entire transparency window of the optical fiber (many 100's of nm at telecommunication wavelengths). The key idea is to pump the SBS process with a beam that is derived from the linearly swept source, but shifted to a higher frequency equal to the Brillouin frequency shift of the fiber $\Omega_B$ using a Mach-Zehnder modulator. In this way, the pump beam frequency automatically tracks the swept-source signal frequency as they enter the fiber and hence are always near the SBS resonance frequency where the slow-light effect is largest. The fact that the pump and signal beams counterpropagate through the fiber causes a small detuning $\Delta v$ between the beams, which decreases the slow light effect. This detuning increases with increasing fiber length $L$ and the source sweep rate $R$ and must be accounted for to optimize the slow light delay.

We experimentally and theoretically investigate the slow light effect and its dependence on $R$ and other experimental parameters. We demonstrate that there is an optimum value of $L$ to obtain the largest delay $\tau$ for a given $R$. We observe $\tau$=10 ns using a 10-m-long photonic crystal fiber (PCF) with $R$=400 MHz/µs and a pump power $P_{in}$=200 mW. Larger delays can be obtained by increasing $P_{in}$ until spontaneous Brillouin scattering dominates the process. We find that the maximum obtainable delay for an optimum-length fiber of the same type used in our experiment is ~38 ns independent of $R$. A pump power of 760 mW is required to obtain the maximum delay for $R$=400 MHz/µs for our fiber.

## 2. Experiment setup

In the experiment, a beam from a single-mode rapidly-tunable laser is split in two, with one beam serving as the SBS pump beam and the other as the signal beam (see Fig. 1). The signal beam passes through a MZM operated in carrier-suppressed mode to create frequency-shifted sidebands at the SBS Stokes and anti-Stokes frequencies. The anti-Stokes beam is filtered out using a fiber Bragg grating (FBG) with a linewidth of 20 GHz. In future research, the MZM and Bragg grating can be removed and replaced by single-side-band carrier-suppressed optical modulator [17]. The pump and signal beams counterpropagate through the optical fiber (the SBS slow light medium), and the signal beam experiences SBS amplification and the associated slow light delay [4]. An additional short fiber is used in the pump beam path to compensate for the propagation time difference between the pump and signal arms. This guarantees the signal and pump have the same frequency shift in the middle of the fiber.

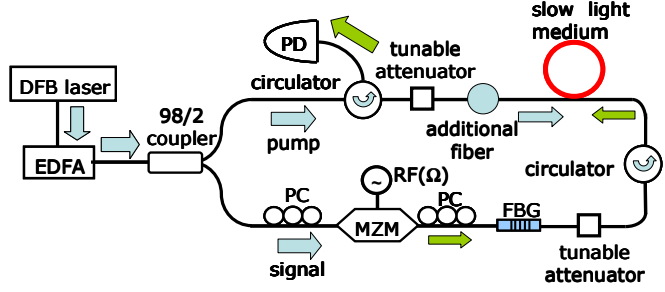

Fig.1. Experiment setup for slow light via stimulated Brillouin scattering. EDFA, erbium doped fiber amplifier; PC, polarization controller; FBG, fiber Bragg grating; PD, photodetector; RF, radio frequency generator.

Before we study the slow light effect with a frequency-swept-source, we first characterize the properties of the fiber for a monochromatic laser ($R=0$). A 10-m-long PCF (NKT Photonics Inc., NL-1550-NEG-1) is used as the slow light medium, which is made of pure silica with a core diameter of 2.1 μm and air-filling fraction of 16%. To measure the gain due to SBS, we record the signal output while turning on and off the pump beam. The gain factor is denoted by $G=\log(P_s/P_{so})$, where $P_s$ ($P_{so}$) is the signal output with (without) the pump beam present, respectively. We measure a Lorentzian-shaped gain profile with a resonance frequency $\Omega_B=9.782$ GHz, and a resonance width (FWHM) $\Gamma_B=40.8$ MHz. To measure $\tau$, we sinusoidally modulate the signal beam intensity at a frequency much less than $\Gamma_B$. We record the transmitted intensity in the presence and absence of the pump beam and determine $\tau$ by comparing the phase difference of the waveforms. We find $\tau=20$ ns with $P_{in}=200$ mW.

### 3. Slow light with a swept source

We investigate the slow light effect using a linearly swept-frequency source. By directly sweeping the injection current of the DFB laser, we obtained a linearly-swept source over a total sweep range of 10 GHz with frequency $\nu(t) = \nu_0 + Rt$, where $\nu_0$ is the initial frequency and $R$ is the sweep rate. The natural linewidth of the DFB laser is ~ 1 MHz, comparable to the linewidth of advanced swept sources used in OCT systems [18]. The repetition period of the modulation waveform is much longer than the transit time of light through the fiber, denoted by $t_r=L/u$, where $u \sim 2\times10^8$ m/s is the speed of light in the fiber, and $t_r$ is ~50 ns for our fiber. Therefore, we can assume the pump and probe frequency increase linearly with time during the SBS process. The MZM is largely insensitive to the change in frequency of the light passing through it. Since the signal and the pump beams are derived from the same swept source, they are chirped simultaneously. The signal-beam frequency automatically tracks the swept-source pump-beam frequency and hence is always near the SBS resonance frequency, which guarantees a strong slow light effect. However, the pump and signal beams counterpropagate through the fiber, which causes a spatially-dependent detuning between the beams and hence decreases the slow light effect.

To investigate the frequency-dependent SBS gain experienced by the chirped beams, we adjust the signal beam frequency (set by the frequency-shift $\Omega$ of the MZM) and measure the gain $G$. Figure 2 shows the measured gain profiles for three different sweep rates as a function of the frequency difference $\delta=\Omega-\Omega_B$. We see that the gain profile is broadened substantially due to the sweep-rate-dependent detuning mentioned above. Broadening begins when the detuning between the two beams differs by $\Omega_B\pm\Gamma_B/2$ at the middle of the fiber (recall that the beams differ by $\Omega_B$ when enter either end of the fiber). That is, when $R=\Gamma_B/(2t_r)\sim400$ MHz/μs for our fiber. To characterize the gain profile in greater detail, we analyze theoretically the gain and delay experienced by the signal beam.

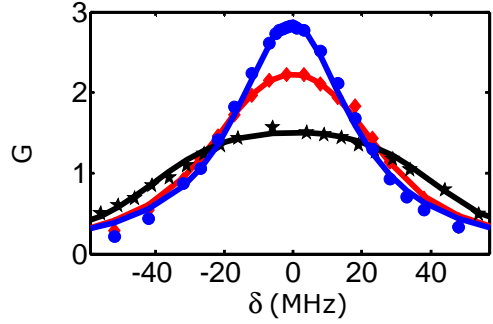

Fig. 2. Measured (dots) and simulated (solid) gain profile of photonic crystal fiber with sweep rate $R$ of 0 (blue circles), 400 MHz/μs (red diamonds), and 800 MHz/μs (black stars).

Consider a reference frame that travels with signal beam (see Fig. 3). The frequency detuning between the pump and signal beam at position $x$ is given by

$$\Delta\nu = \nu_s - \nu_p(x) + \Omega_B = \nu\left(t - \frac{x}{u}\right) - \nu\left(t - \frac{L-x}{u}\right) + \delta = \frac{L-2x}{L}Rt_r + \delta, \quad (1)$$

where $\nu_s$ and $\nu_p(x)$ are frequencies of signal and pump beams, respectively. The detuning $\Delta\nu$ decreases $G$, whose effect can be found by integrating the differential gain per unit path length $dx$ over the whole fiber as

$$G = \ln\left(\frac{P_s}{P_{so}}\right) = \int_0^L g(\Delta\nu)P(x)dx, \quad (2)$$

where $g(\Delta\nu)$ is gain profile of the slow light medium, and $P(x)$ is pump power. We find

$$G = \int_0^L g_p P_{in} e^{-\alpha x} \frac{\left(\frac{\Gamma_B}{2}\right)^2}{\left(\frac{L-2x}{L}Rt_r + \delta\right)^2 + \left(\frac{\Gamma_B}{2}\right)^2} dx, \quad (3)$$

where $\alpha$ ($g_p$) is the optical loss (SBS gain) coefficient of the fiber. For our photonic crystal fiber, $\alpha=9$ dB/km and $g_p=2.5$ m$^{-1}$W$^{-1}$. The PCF used in this experiment is only 10 meter long, which makes the optical loss negligible and hence we can set $\alpha=0$ for simplicity. However, for a longer fiber, the optical loss needs to be considered because it will cause asymmetry of the broadened gain profile. With $\alpha=0$, we solve Eq. (3) and find

$$G = g_p P_{in} \Gamma_B L \frac{\arctan\left(\frac{2t_r R + 2\delta}{\Gamma_B}\right) + \arctan\left(\frac{2t_r R - 2\delta}{\Gamma_B}\right)}{2t_r R}. \quad (4)$$

The linewidth (FWHM) of the gain profile is then given by

$$\Gamma = \sqrt{\Gamma_B^2 + 4t_r^2 R^2}, \quad (5)$$

and goes over to $2t_rR$ in the limit of $R>>\Gamma_B/(2t_r)$. Using Eq. (4), we predict the SBS gain profile for three different sweep rates, as shown in Fig. 2. Agreement with our experimental measurements is very good.

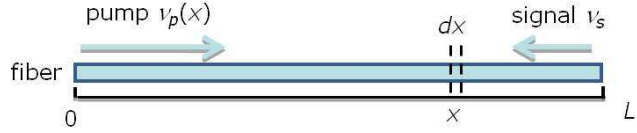

Fig. 3. Diagram of the space-dependent frequency detuning between the pump and signal beams caused by the swept source.

The slow light effect is largest at the center of the resonance ($\Omega=\Omega_B$). Therefore, we analyze the gain and delay and their dependence on $R$ on the resonance. With $\delta=0$ in Eq. (4), we find that

$$G = g_p P_{in} \Gamma_B L \frac{\arctan\left(\frac{2t_r R}{\Gamma_B}\right)}{2t_r R}. \qquad (6)$$

The slow light delay due to the SBS process is given by

$$\tau = \int_0^L \frac{n_g(x) - n_{fg}}{c} dx, \qquad (7)$$

where $n_{fg}$ is the group index of the fiber without the SBS process, $c$ is speed of light in vacuum, and

$$n_g(x) = n_{fg} + \frac{cg_p P_{in}}{2\pi\Gamma_B} \frac{1 - \frac{4\Delta\nu^2}{\Gamma_B^2}}{\left(1 + \frac{4\Delta\nu^2}{\Gamma_B^2}\right)^2} \qquad (8)$$

is the group index due to the SBS effect in a differential segment of the fiber [6]. By combining Eqs. (1), (7) and (8), we find

$$\tau = \frac{g_p P_{in} L \Gamma_B}{2\pi\Gamma^2}. \qquad (9)$$

The gain and delay are measured in our PCF using $P_{in}$=200 mW for various value of $R$ as shown in Fig.4, using the same method described in the previous section. We see that the delay drops faster than the gain due to the broadened gain profile, as discussed above. In particular, we observe $\tau$=10 ns for $R$=400 MHz/μs and $\tau$=4 ns for $R$=800 MHz/μs. We also overlay the predictions of Eq. (6) and (9) with no free parameters, where the agreement with our observations is very good.

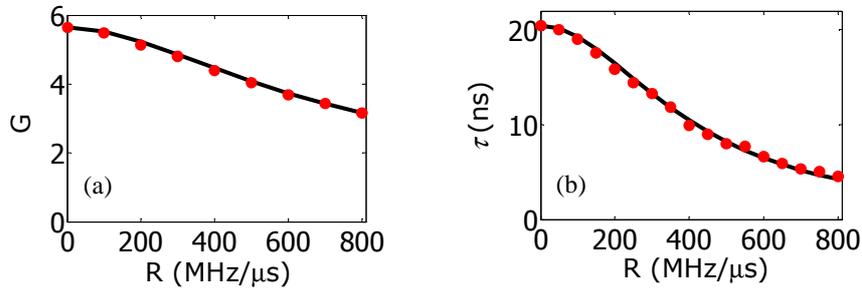

Fig. 4. (a) Measured (dots) and simulated (solid) gain as a function of sweep rate $R$. (b) Measured (dots) and simulated (solid) delay as a function of sweep rate $R$.

## 4. Optimum fiber length to achieve the largest delay

We find that, for a given sweep rate $R$, there is an optimum value of fiber length $L$ to obtain the largest delay. From Eqs. (5) and (9), the maximum delay occurs for $L_{opt}= u\Gamma_B/(2R)$ and is equal to $\tau_{max}=g_pP_{in}u/(4R)$, showing that $\tau_{max}$ is inversely proportional to $R$. Figure 5(a) shows the delay as a function of $L$ for $R=400$ MHz/μs using Eq. (9). For $\Gamma_B=40$ MHz, $L_{opt}\sim10$ m, which corresponds to the length of our PCF. For fixed $L$ and different $R$, a constant delay can be obtained by adjusting $P_{in}$, as shown in Fig. 5(b).

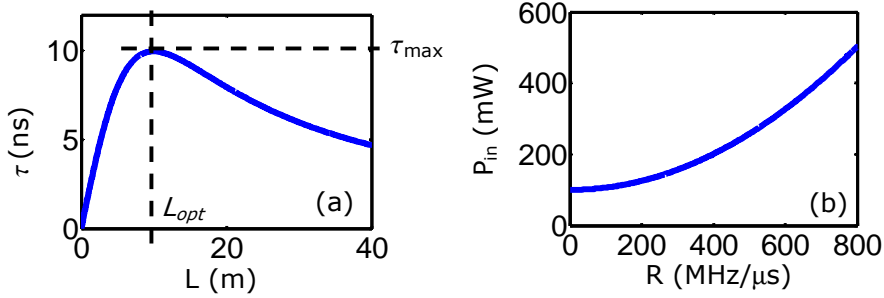

Fig. 5. (a) Delay as a function of $L$ for $R=400$ MHz/μs with $P_{in}=200$ mW. (b) The pump power needed to obtain $\tau=10$ ns for $L=10$ m.

Larger delays can be obtained by increasing the pump power until spontaneous Brillouin scattering dominates the process. The gain parameter $G$ is limited up to the Brillouin threshold $G_{th}$, where the delay reaches its maximum value $\tau_{th}$ obtainable by the fiber. By combining Eqs. (5) and (8) and assuming $L=L_{opt}$ (giving $Rt_r=\Gamma_B/2$), we find that

$$\tau_{th} = \frac{G_{th}}{\pi^2\Gamma_B}, \qquad (10)$$

which is independent of $R$. Using $\Gamma_B=40$ MHz and $G_{th}=15$ for the 10-m-long PCF [19], we estimate $\tau_{th}=38$ ns, obtained when $P_{in}=760$ mW.

## 5. Conclusion

We investigate the slow light effect via SBS with a linearly swept-frequency source. The pump and signal beams counterpropagate through the fiber, which introduces a small detuning between the beams and hence decreases the slow light effect. This detuning increases with increasing fiber length $L$ and the source sweep rate $R$. We find there is an optimum value of fiber length to obtain the largest delay for a given sweep rate. Using the optimum-length fiber, we observed a delay of 10 ns with $R$ of 400 MHz/μs and a pump power of 760 mW. The slow light can be achieved, in principle, over the entire transparency window of the optical fiber if we were to use a laser with a larger sweep range and a single-side-band carrier-suppressed modulator. It has a potential to work with the high-speed commercial frequency-swept sources, which makes it applicable to optical coherence tomography and Fourier transform spectroscopy.

**Acknowledgment**

R.Z., Y.Z. and D.J.G. gratefully acknowledge the financial support of the DARPA DSO Slow Light Program and the Air Force Research Laboratory under contract FA8650-09-C-7932. J.W. gratefully acknowledges the financial support of the China Scholarship Council and Beijing Jiaotong University.